# MISTIQS: An open-source software for performing quantum dynamics simulations on quantum computers

Connor Powers[1], Lindsay Bassman[1,2], Thomas M. Linker[1], Ken-ichi Nomura[1], Sahil Gulania[3], Rajiv K. Kalia[1], Aiichiro Nakano[1] and Priya Vashishta[1]

[1] *Collaboratory for Advanced Computing and Simulations, University of Southern California, Los Angeles, CA 90089-0242, USA*

[2] *Lawrence Berkeley National Laboratory, Berkeley, CA 94720, USA*

[3] *Department of Chemistry, University of Southern California, Los Angeles, CA 90089-1062, USA*

**Abstract.**
We present *MISTIQS*, a Multiplatform Software for Time-dependent Quantum Simulations. *MISTIQS* delivers end-to-end functionality for simulating the quantum many-body dynamics of systems governed by time-dependent Heisenberg Hamiltonians across multiple quantum computing platforms. It provides high-level programming functionality for generating intermediate representations of quantum circuits which can be translated into a variety of industry-standard representations. Furthermore, it offers a selection of circuit compilation and optimization methods and facilitates execution of the quantum circuits on currently available cloud-based quantum computing backends. *MISTIQS* serves as an accessible and highly flexible research and education platform, allowing a broader community of scientists and students to perform quantum many-body dynamics simulations on current quantum computers.

## 1. Motivation and Significance

With Google's recent experimental realization of quantum supremacy for a proof-of-concept problem [1] and IBM's announced roadmap for scaling quantum computers up to over a thousand qubits as early as 2023, there is a growing demand for the use of quantum computers for nontrivial scientific applications. A highly anticipated application is as a universal simulator of quantum many-body systems, an idea originally conceived of by Richard Feynman in the 1980s [2] and later elaborated by Seth Lloyd [3]. The last decade has witnessed the growing success of quantum computing for simulating *static* properties of quantum systems, *i.e.*, the ground state energy of *small* molecules [4-9]. However, it remains a challenge to simulate quantum many-body *dynamics* on current-to-near-future noisy intermediate-scale quantum (NISQ) computers [10].

To facilitate the adoption of quantum computing for studying quantum dynamics, we have developed open-source software to perform non-trivial many-body quantum dynamics on the publicly available IBM-Q and Rigetti quantum computers. While there are general open source platforms such as *Cirq* [11] for developing general quantum circuits to be run on quantum computers, a host of emergent quantum programming languages [20], and other problem-specific platforms such as *OpenFermion* [12] for solving static electronic structure problems on quantum computers, no such platform is widely available for studying time-dependent quantum many-body dynamics. Here, we present MISTIQS (Multiplatform Software for Time-dependent Quantum Simulation), an open-source software package dedicated to enabling the dynamic simulation of quantum many-body systems that can be represented by the Heisenberg model, a ubiquitous model that captures the behavior of a variety of quantum materials and systems. An early prototype of this software was successfully used to simulate ultrafast control of emergent magnetism by terahertz radiation in atomically thin Re-doped $MoSe_2$ monolayers [13], and has since been expanded to simulate the dynamics of broader material systems that are described by the Heisenberg model.

In addition, MISTIQS also includes a domain-specific quantum circuit compiler for simulating a subgroup of the Heisenberg model known as the transverse field Ising model (TFIM) [22]. By lowering the gate count of the circuits for simulating the many-body dynamics, this special-purpose compiler allows for more accurate, longer-time simulations by reducing the compounding gate error [14], while also significantly reducing the wall-clock compilation time over backend-native general-purpose compilers. In total, MISTIQS employs a user-friendly, object-orientated framework for formulating, optimizing, and executing quantum circuits for dynamic many-body simulations on quantum computers with the goal of expanding education and spurring the development of research in this field.

## 2. Software Description

*MISTIQS* is written in Python, with backend-specific libraries only imported as needed per each use case. *MISTIQS* provides a full-stack solution for the direct quantum simulation of spin systems governed by the Heisenberg model Hamiltonian, taking the following form for *N* spins:

$$H(t) = -\sum_{i=1}^{N-1}[J_x\sigma_i^x\sigma_{i+1}^x + J_y\sigma_i^y\sigma_{i+1}^y + J_z\sigma_i^z\sigma_{i+1}^z] - h(t)\sum_{i=1}^{N}\sigma_i^k \quad k \in \{x,y,z\} \quad 1$$

Here, $J_x$, $J_y$, and $J_z$ give the exchange interaction strengths between nearest neighbor spins in the $x$-, $y$-, and $z$-directions respectively, $h(t)$ gives the time-dependent external magnetic field to which the spins are exposed, and $\sigma_i^k$ is the $k$-th Pauli matrix acting on qubit $i$. The dynamics of systems modeled by this class of Hamiltonians can be simulated on digital quantum computers by mapping the states of the spins to those of the qubits, and translating the Hamiltonian-dependent time-evolution operator into a quantum circuit. Execution of the resultant quantum circuits on a quantum computer produces results that can be post-processed to show the dynamic evolution of the system.

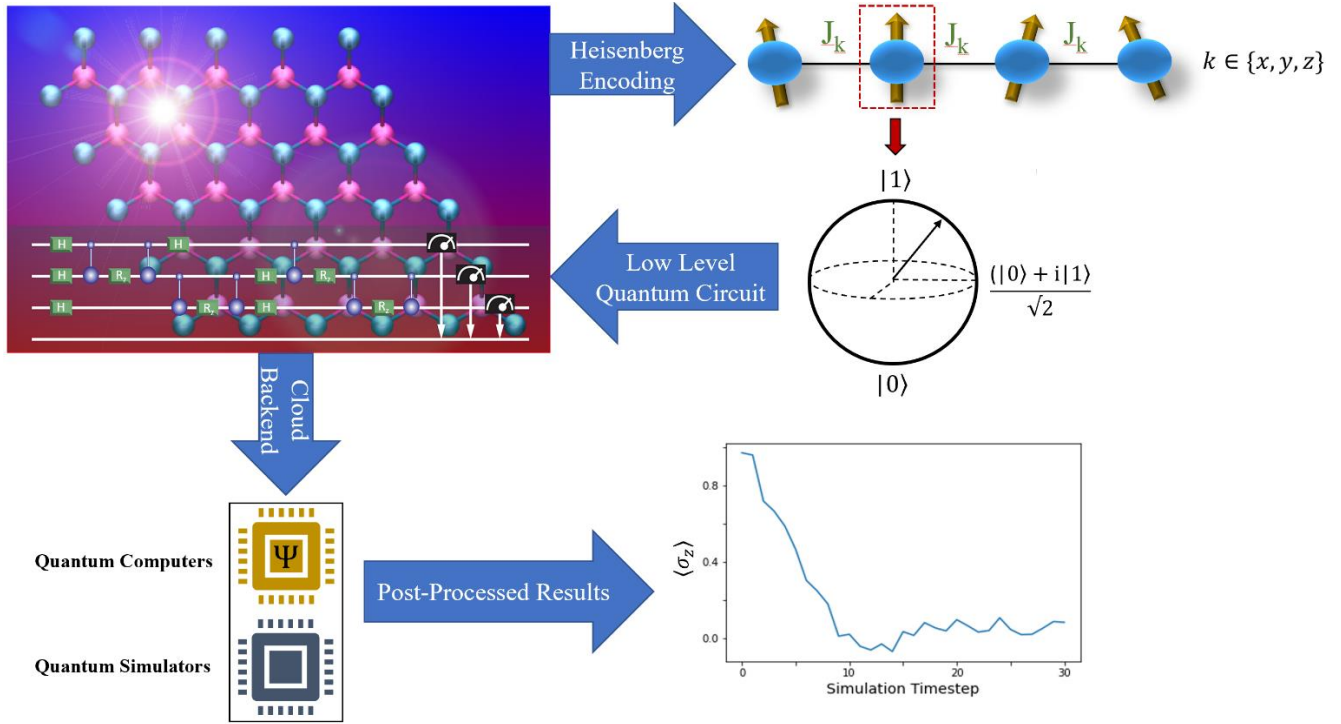

**Fig. 1.** Process overview of simulating Heisenberg spin Hamiltonians on digital quantum computers and quantum simulators.

The workflow for MISTIQS is described by Fig. 1. First, the user must define the coefficients of the Hamiltonian specific to the material system of interest, along with other simulation parameters such as number of time-steps, step-size, and quantum backend. *MISTIQS* then generates quantum circuits that simulate the time-evolution of the system. These circuits can then be executed on the chosen NISQ backend. Finally, the results can be post-processed to reveal the evolution of the spins over the course of the simulation. It is worth emphasizing that *MISTIQS* accepts a user-specified function for the time dependence of the external magnetic field, defined as $h(t)$ in Equation 1, allowing researchers to more accurately model the specific conditions of their experiments. The specifics of this full-stack functionality are shown in more detail in **Fig. 2**a.

The user can also choose to only use portions of *MISTIQS*'s functionality. For example, if the user just wants to output a list of quantum circuits compatible on IBM quantum devices, they can choose to skip circuit execution and post-processing. There is also the option to use the domain-specific compilers on externally generated circuits to output compiled circuits for IBM's or Rigetti's platforms.

*2.1. Software Architecture*

Upon downloading the software from https://github.com/USCCACS/MISTIQS, the following directories will be present:

- **src/:** Directory containing *MISTIQS* source code.
- **docs/:** Directory containing user's manual for *MISTIQS*.
- **examples/:** Directory containing demonstrative examples covering some different use cases of *MISTIQS*.

Upon running any portion of the software, the following subdirectory will be created:

- **data/:** Directory containing simulation results, graphics (if applicable), and the logfile generated by executing the software.

*MISTIQS* is comprised of three core modules, described below with additional information about their key member functions:

- **quantum_circuits:** Code defining quantum logic gate and quantum circuit objects native to *MISTIQS* that allow it to operate above the syntax of any one quantum computing platform.
    - **Gate**: Takes as input a quantum gate name, rotation angles (if applicable), and the set of qubits it acts on to create an intermediate quantum gate representation.
    - **Program**: Builds an intermediate quantum circuit representation from a list of **Gate** objects.
- **Heisenberg:** Code that generates, compiles, and executes quantum circuits using user-defined Hamiltonian and simulation as input.
    - **generate_circuits**: From the input file, this method uses the user-specified Hamiltonian parameters, as well as user choices regarding backend and compilation. First, it generates intermediate quantum circuits needed to simulate the time-evolution of the system, then it compiles these circuits into the native gate sets and syntax of the user-specified quantum computing backend.
    - **connect_ibm**: For IBM use cases, this method connects to the IBMQ backend (needed for quantum circuit compilation and execution). Takes in IBMQ API key and account overwrite boolean as needed.
    - **run_circuits**: From the input file, this method uses the backend and quantum device choice, as well as user choices regarding post-processing of results. Its functionality is to execute the quantum circuits on the user-specified quantum device and post-process the results to the user's specifications.
- **ds_compiler:** Code for domain-specific compilation of circuits for the TFIM (which is a special case of the general Hamiltonian in Equation 1) into the native gate sets employed by IBM and Rigetti.
    - **ds_compile_ibm:** Domain-specific quantum compiler for circuits simulating time-evolution of the TFIM. Takes in a high-level quantum circuit and returns a compiled quantum circuit executable on IBM quantum computers.

- **ds_compile_rigetti:** Domain-specific quantum compiler for circuits simulating time-evolution of the TFIM. Takes in a high-level quantum circuit and returns a compiled quantum circuit executable on Rigetti quantum computers.

The described roles and key member functions of these modules are illustrated in **Fig. 2**b. In this figure, information inlets and outlet paths are also highlighted, and the optional compiler-only use case is included in the context of the nominal workflow.

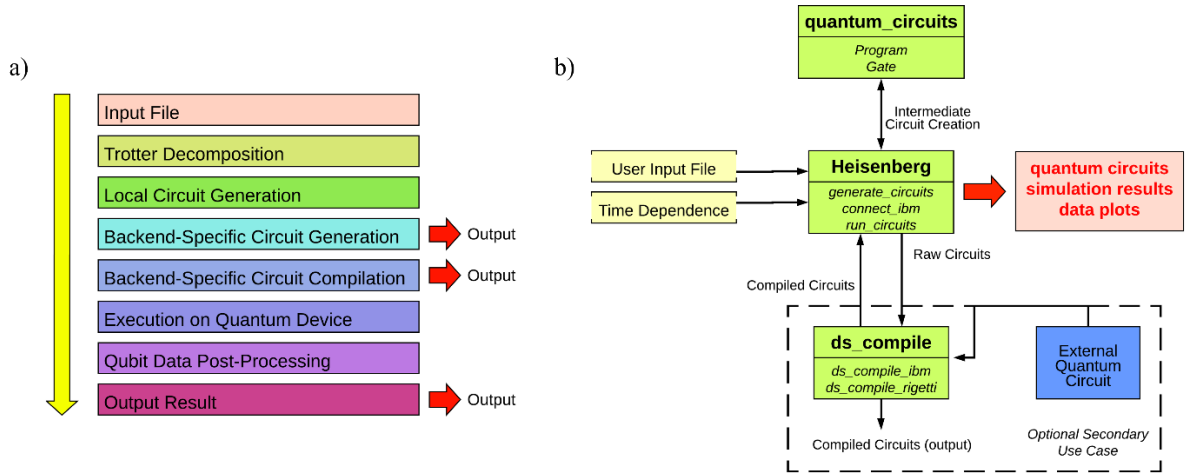

*Fig. 2. a) Full-stack representation of MISTIQS workflow. b) Key roles and member functions of MISTIQS modules.*

This multi-output architecture allows *MISTIQS* to serve a variety of education-facing and research-facing applications; utilizing the full feature stack allows for streamlined visualizations of key observable evolutions under the Hamiltonian of interest, while the ability to output the generated simulation circuits allows for researchers to utilize their own compilers or transform circuits into the syntax appropriate for alternate quantum hardware backends.

## 2.2. Prerequisites

The *MISTIQS* codebase was intentionally written to minimize the number of required external libraries; it only requires the *numpy* library to be installed. If the user would like to generate plots of the simulation results, then the *matplotlib* library must be installed as well. Otherwise, any additional required libraries are solely dependent on the quantum-computing platform the user would like to produce quantum circuits for; qiskit [17] is required to generate circuits for IBM devices, pyquil [18, 19] is required for Rigetti use cases, and cirq [11] is likewise required for Google use cases.

## 2.3. Generalized Workflow

To use *MISTIQS*, the user must first define any parameters that need to be changed from their default values in a simple text-based input file. The customizable parameters are described in the appendix. A *Heisenberg* object is then created, which stores all of the input system parameters. Note that while attributes of the object are initially set from the input file, they can be changed later in the code as any time. Next, the software transforms the input Hamiltonian and simulation parameters into a series of quantum programs (an intermediate representation of quantum circuits) that will simulate the dynamics of the system of interest via Trotter approximation. This intermediate circuit representation (native to *MISTIQS*) facilitates the transformation of the quantum circuits into the equivalent native circuits for various quantum computing platforms. This is crucial because quantum computing platforms can vary widely in not only in circuit object syntax, but also the native quantum gate set allowed on their hardware.

*MISTIQS* then compiles the circuits, if desired, using either the compilers native to the specified quantum computing platform or the built-in domain-specific quantum compilers. If execution of the quantum simulation is desired, *MISTIQS* will run the circuits on the specified quantum device. Once successfully run, the software can post-process the results and save the average magnetization data for each qubit over the duration of the simulation in individual output files. It can also generate and save plots of this data. A logfile is kept during each use, and this, along with all generated qubit data and plots, will be saved in a **data** folder.

*2.4. Software Functionalities*

*MISTIQS* can perform quantum dynamics simulations for any variant of the time-dependent Heisenberg model, including the important XX chain model, XXZ chain model, and TFIM [21]. The time-dependent external magnetic field, set to a sinusoidal function of a tunable frequency by default, can also be customized by the user, allowing for greater flexibility in the pulse shapes that can be simulated by this software. Another key functionality of the software is quantum circuit compilation, as its built-in domain-specific compilers are optimized for TFIM simulations on IBM and Rigetti backends.

**3. Illustrative Examples**

*3.1. XX Chain Model Domain Wall Quench Simulation*

In this example, we use a 16-qubit noiseless quantum simulator to explore the dynamics of a 6-spin XX chain forming a domain wall. We are interested in measuring and plotting the average magnetization of each spin over time. An XX spin chain is a subcase of Equation 1 where $J_x = J_y \neq 0$ and $J_z = h(t) = 0$, with $J_x$ set by the material being simulated. While it is one of the simplest subsets of the Heisenberg model, it has shown relevance in entanglement teleportation and low-energy quantum chromodynamics (QCD) [15,16].

For this example, we configure the initial spins of the spin chain by setting the *initial_spins* parameter so that half of the spins start in a spin-down configuration in order to form a domain wall. To run the simulation, we first initialize the *Heisenberg* object with the relevant input file, then generate the circuits with the *generate_circuits()* method, then run the simulation with the *run_simulation()* method. Fig. 3a illustrates the spin states of each spin in this domain wall example over the course of a 50 fs simulation, with the evolution of the second spin highlighted in green, and the corresponding plot of the evolution of the second spin produced by *MISTIQS* is shown in Fig. 3b. The results align with those presented in Fig 2 of ref [21].

*3.2. Quantum Ising model Simulation of Emergent Magnetism in MoSe$_2$:*

In this example, we will explore the emergent magnetism of a Re-doped MoSe$_2$ monolayer, which we model with the TFIM, by performing a 5-qubit quantum simulation on IBM's 5-qubit "Ourense" quantum computer. TFIM spin chains are governed by Equation 1 where $J_x = J_y = 0$, $J_z \neq 0$, and $k = x$. Physically, this describes a system with inter-spin coupling only in the $z$-direction in the presence of an external magnetic field aligned with the $x$-direction. To run a quantum simulation of such a system with *MISTIQS*, set the *Jz* parameter to the value matching the system of interest, set the *ext_dir* parameter to '*X*', then set the *h_ext* to a nonzero value. Then, the *num_qubits* parameter is set accordingly, and the *initial_spins* parameter is set to reflect standard spin-up configuration. Other general simulation parameters, such as simulation length and backend, are also set. A complete example input file, **TFIM_input_file.txt**, is found in the **examples** directory.

To run the simulation, a *Heisenberg* object is created with the relevant input file, then the user is connected to the IBM Q backend by running the *connect_IBM()* method. Next, the quantum circuits are generated by running the *generate_circuits()* method, and the simulation is executed and post-processed by running the *run_circuits()* method. The resulting plot of average magnetization should resemble Fig. 3c, although differences will inevitably arise due to device noise.

*3.3. Standalone Compiler for the Quantum Ising Problem:*

In this example, we will use the domain-specific (DS) compilers built into *MISTIQS* to optimize the quantum circuits of a TFIM simulation. The user need only import the ds_compiler module, and call the *ds_compile* method on the circuits, providing the desired backend to compile to as an argument. If running the sample circuits described above, this backend argument would be "ibm". The optimized circuits will be returned as a new list. Input circuits may be generated by initializing a *Heisenberg* object with the TFIM example input file (see Section 3.2), and using the output of the *return_circuits()* method of the *Heisenberg* object. Alternatively, the user may input externally created circuits.

The **examples** directory contains an example use case that directly implements these steps, then runs the same circuits through IBM's native compiler to directly compare the performances of the two compilers. This will produce a comparison of quantum gate counts between the identical circuits compiled by the domain-specific compiler built into *MISTIQS* and the native IBM compiler resembling **Fig. *3***d.

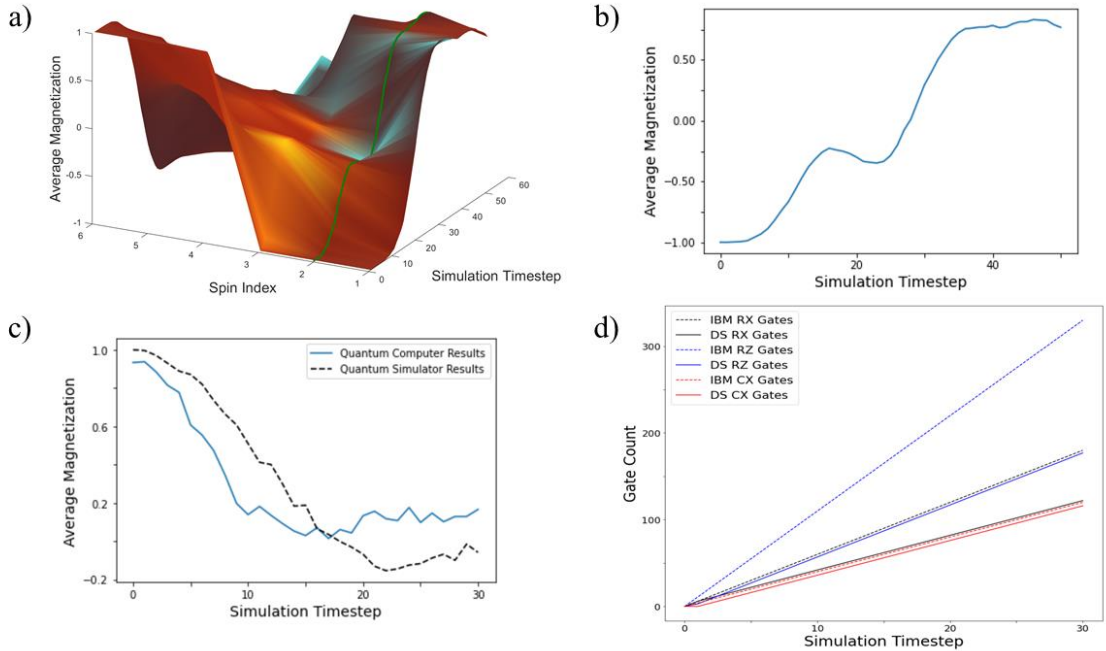

**Fig. 3.** (a) Average magnetization of the domain wall example spin chain over time. (b) Average magnetization of the second spin over time as output by MISTIQS. (c) Example average magnetization evolution of a TFIM chain spin executed by MISTIQS on both a noiseless quantum simulator and an IBM quantum computer. (d) Comparison of quantum gate counts in a sample TFIM circuit when compiled with the IBM native compiler and the domain-specific compiler integrated into MISTIQS.

## 4. Impact

The three major impacts of *MISTIQS* are: (1) to provide an open-source platform enabling accessible, end-to-end quantum dynamics simulations across the major quantum computing platforms open to public use; (2) to accelerate quantum dynamics simulation research by wrapping quantum circuit formation, compilation, execution, and basic post-processing into a user-friendly package; and (3) to facilitate quantum computing education, where the accompanying examples and tutorials can be used as self-contained course modules in classroom and workshop settings.

*MISTIQS* aims to bridge the gap between quantum computation and fields such as chemistry, materials science, and condensed matter physics by making quantum dynamics simulations on NISQ-era devices accessible to a wider range of researchers. It is our hope that this high-level programming library can spur new acceleration in exploring the power of current quantum devices across a wide range of applications. Already, beta versions of this software have been successfully used to perform quantum simulations of ultrathin materials under the TFIM exploring emergent magnetism phenomena [13].

## 5. Conclusions

In summary, *MISTIQS* is an end-to-end software solution for performing quantum simulations of systems governed by any subset of the time-dependent Heisenberg model Hamiltonians on current quantum computers. By simplifying and streamlining the workflow from the material's Hamiltonian to the post-processed quantum computer data, as well as providing effortless cross-platform functionality between IBM, Google, and Rigetti quantum computing platforms, researchers outside of the field of quantum computation can easily leverage the power of quantum computers in their work. *MISTIQS* has already proven its utility in the simulation of quantum materials, and is expected to accelerate quantum simulation research in a wide variety of applications including quantum chemistry, materials science, and condensed matter physics.

## Acknowledgements

This work was supported as part of the Computational Materials Sciences Program funded by the U.S. Department of Energy, Office of Science, Basic Energy Sciences, under Award Number DE-SC0014607.

**Appendix**

List and Descriptions of Customizable Parameters

| Parameter Name | Description |
| --- | --- |
| **Jx** | Controls inter-spin coupling in the X direction |
| **Jy** | Controls inter-spin coupling in the Y direction |
| **Jz** | Controls inter-spin coupling in the Z direction |
| **h_ext** | Controls amplitude of the Hamiltonian's external magnetic field term |
| **ext_dir** | Sets the direction of the external magnetic field |
| **num_qubits** | Specifies the number of qubits of the quantum simulation |
| **initial_spins** | Sets the initial spins of each qubit |
| **delta_t** | Sets the timestep of the quantum simulation |
| **steps** | Specifies the number of timesteps in the quantum simulation |
| **QCQS** | Specifies whether the circuits will be run on a quantum computer or a quantum simulator |
| **shots** | Sets the number of shots to execute on the quantum device |
| **noise_choice** | If using a quantum simulator, this sets whether a noisy simulator is desired. |
| **device_choice** | Specify the quantum device to generate the circuits for and\or run the circuits on |
| **plot_flag** | Sets whether post-processed results from running the quantum simulation should be plotted and saved. |
| **time_dep_flag** | Specifies whether time dependence is desired in the Hamiltonian external field term |
| **freq** | Specifies the frequency of the optional time dependent function in the Hamiltonian external field term |
| **custom_time_dep** | Sets whether the software should look for a user-defined time dependence function for the Hamiltonian external field term |
| **backend** | Sets the choice of quantum computing platform (IBM, Rigetti, Google) |
| **compile** | Sets whether the software should compile the generated quantum circuits |
| **auto_smart_compile** | Sets whether the integrated domain-specific compilers should automatically be applied to detected TFIM circuits (IBM and Rigetti backends only) |
| **default_compiler** | Sets whether the software defaults to compilers native to the backend of choice or the integrated domain-specific compilers (IBM and Rigetti backends only) |

## B- Required Metadata

### B1 Current executable software version

*Table 1 – Software metadata*

| Nr | (executable) Software metadata description | *Please fill in this column* |
|---|---|---|
| | | |

| S1 | Current software version | *v1.0* |
|---|---|---|
| S2 | Permanent link to executables of this version | *https://github.com/USCCACS/MISTIQS* |
| S3 | Legal Software License | *MIT License* |
| S4 | Computing platform / Operating System | *Linux* |
| S5 | Installation requirements & dependencies | *Python 3.8, numpy, matplotlib\*, qiskit\*, pyquil\*, cirq\** <br> *\*optional depending on use case* |
| S6 | If available Link to user manual - if formally published include a reference to the publication in the reference list | *https://github.com/USCCACS/MISTIQS/tree/master/docs* |
| S6 | Support email for questions | *cdpowers@usc.edu* |

## B2 Current code version

*Table 2 – Code metadata*

| Nr | Code metadata description | *Please fill in this column* |
|---|---|---|
| C1 | Current Code version | *v1.0* |
| C2 | Permanent link to code / repository used of this code version | *https://github.com/USCCACS/MISTIQS* |
| C3 | Legal Code License | *MIT License* |
| C4 | Code Versioning system used | *git* |
| C5 | Software Code Language used | *python* |
| C6 | Compilation requirements, Operating environments & dependencies | *Python 3.8, numpy, glob\*, matplotlib\*, qiskit\*, pyquil\*, cirq\** <br> *\*optional depending on use case* |
| C7 | If available Link to developer documentation / manual | *https://github.com/USCCACS/MISTIQS/tree/master/docs* |
| C8 | Support email for questions | *cdpowers@usc.edu* |